\documentclass[twocolumn,preprintnumbers,prd,nofootinbib,superscriptaddress]{revtex4-1}

\usepackage{amsmath}
\usepackage{amssymb}
\usepackage{amsthm}
\usepackage{graphicx}
\usepackage{subfigure}
\usepackage{changepage}
\usepackage[breaklinks]{hyperref}
\usepackage{url}
\usepackage{breakurl}
\usepackage{mathtools,xparse}
\usepackage{verbatim}
\usepackage{dsfont}

\theoremstyle{definition}

\newtheorem*{definition*}{Definition}

\def\app#1#2{%
  \mathrel{%
    \setbox0=\hbox{$#1\sim$}%
    \setbox2=\hbox{%
      \rlap{\hbox{$#1\propto$}}%
      \lower1.1\ht0\box0%
    }%
    \raise0.25\ht2\box2%
  }%
}

\DeclarePairedDelimiter{\norm}{\lVert}{\rVert}

\begin{document}

\title{Comments on Holographic Entanglement Entropy in {\boldmath $TT$} Deformed CFTs}

\author{Chitraang Murdia}
\affiliation{Berkeley Center for Theoretical Physics, Department of Physics, 
  University of California, Berkeley, CA 94720, USA}
\affiliation{Theoretical Physics Group, Lawrence Berkeley National Laboratory, 
  Berkeley, CA 94720, USA}
\author{Yasunori Nomura}
\affiliation{Berkeley Center for Theoretical Physics, Department of Physics, 
  University of California, Berkeley, CA 94720, USA}
\affiliation{Theoretical Physics Group, Lawrence Berkeley National Laboratory, 
  Berkeley, CA 94720, USA}
\affiliation{Kavli Institute for the Physics and Mathematics 
 of the Universe (WPI), The University of Tokyo Institutes for Advanced Study, 
 Kashiwa 277-8583, Japan}
\author{Pratik Rath}
\affiliation{Berkeley Center for Theoretical Physics, Department of Physics, 
  University of California, Berkeley, CA 94720, USA}
\affiliation{Theoretical Physics Group, Lawrence Berkeley National Laboratory, 
  Berkeley, CA 94720, USA}
\author{Nico Salzetta}
\affiliation{Berkeley Center for Theoretical Physics, Department of Physics, 
  University of California, Berkeley, CA 94720, USA}
\affiliation{Theoretical Physics Group, Lawrence Berkeley National Laboratory, 
  Berkeley, CA 94720, USA}

\begin{abstract}
The Ryu-Takayanagi (RT) formula has been a key ingredient in our understanding of holography. Recent work on $TT$ deformations has also boosted our understanding of holography away from the conformal boundary of AdS. In this short note, we aim to refine some recent work demonstrating the success of the RT formula in $TT$ deformed theories. We emphasize general arguments that justify the use of the RT formula in general holographic theories that obey a GKPW-like dictionary. In doing so, we clarify subtleties related to holographic counterterms and discuss the implications for holography in general spacetimes.
\end{abstract}

\maketitle

\makeatletter
\def\l@subsection#1#2{}
\def\l@subsubsection#1#2{}
\makeatother

\section{Introduction}
\label{sec:intro}

Gauge-gravity duality, specifically AdS/CFT, is our best known example of a holographic description of quantum gravity \cite{Maldacena:1997re}. The so-called GKPW dictionary \cite{Gubser:1998bc,Witten:1998qj} relating bulk physics to  boundary dynamics takes the form
\begin{align}
 Z_{\text{CFT}}[\gamma_{ij}] = e^{-I_{\text{bulk}}[g_{\mu\nu}]},
\label{eq:dict}
\end{align}
where $\gamma_{ij}$ is the background metric of the space in which the boundary CFT lives, and $g_{\mu\nu}$ is the bulk metric. A particularly consequential holographic correspondence given by this duality is the Ryu-Takayanagi (RT) formula
\begin{align}
 S(A) = \min_{\partial \Gamma = \partial A} \left[\frac{\norm{\Gamma}}{4G}\right],
\label{eq:RT}
\end{align}
which relates the entanglement entropy of a subregion $A$ of the boundary space to the area of the bulk extremal surface $\Gamma$ anchored to the entangling surface $\partial A$ \cite{Ryu:2006bv,Ryu:2006ef,Hubeny:2007xt}. Throughout, we will work to leading order in the bulk Newton's constant, $G$, and suppress all bulk fields aside from $g_{\mu\nu}$. Higher order effects are well understood in the context of AdS/CFT \cite{Faulkner:2013ana,Engelhardt:2014gca}.

Other holographic dualities with similar features have been proposed. In particular, the $TT$ deformation of 2-dimensional CFTs and its appropriate generalizations to higher dimensions have been argued to have holographic duals~\cite{McGough:2016lol,Taylor:2018xcy,Hartman:2018tkw}. Of crucial importance to our discussion is that the proposed dictionary relating the boundary and bulk observables in these theories takes the same form as Eq.~(\ref{eq:dict}), except that now Dirichlet boundary conditions are imposed on a cutoff surface in the bulk.

The simple idea we would like to highlight is that Eq.~(\ref{eq:RT}) was derived in the context of AdS/CFT from Eq.~(\ref{eq:dict}) under rather tame assumptions \cite{Lewkowycz:2013nqa}. The same argument, therefore, can be used to show that the RT formula holds for all dualities adopting dictionaries of the form of Eq.~(\ref{eq:dict}). This straightforward result is known in the community; however, careful consideration of it resolves subtleties involving counterterms when calculating entanglement entropy in $TT$ deformed theories.

We start by reviewing some aspects of entanglement entropy from a field theory perspective in Section~\ref{sec:QFTbasic}. We then proceed to a calculation in the particular case of $TT$ deformed theories in Section~\ref{sec:QFTcalc}. We discuss the general holographic argument for the RT formula in Section~\ref{sec:bulkbasic}, which is followed by a sample calculation in cutoff AdS in Section~\ref{sec:bulkTT}. Along the way, we address some subtleties related to holographic renormalization. We conclude with a discussion about the consequences for holography in general spacetimes in Section~\ref{sec:disc}.

Note that several calculations of entanglement entropy in $TT$ deformed theories have appeared recently \cite{Chakraborty:2018kpr,Donnelly:2018bef,Chen:2018eqk,Gorbenko:2018oov,Park:2018snf,Caputa:2019pam,Banerjee:2019ewu}. Our goal is to emphasize the generality of the arguments leading to an agreement between boundary entanglement entropy and the RT formula and clarify some of the calculations performed in these works.

\subsection*{Conventions}
The background metric of the space in which the boundary field theory lives is denoted by $\gamma_{ij}$, while $h_{ij}$ refers to the bulk induced metric on the cutoff surface at $r=r_{c}$. These are related by $h_{ij}=r_{c}^{2}\gamma_{ij}$.

\section{Field Theory Calculation}
\label{sec:FT}

\subsection{Preliminaries}
\label{sec:QFTbasic}

Consider a $D$-dimensional CFT with action $I[\phi] = \int d^{D}\!x \sqrt{\gamma}\, \mathcal{L}[\phi]$. One can prepare a density matrix $\rho$ on a spatial slice $\Sigma$ using an appropriate Euclidean path integral. In order to compute the entanglement entropy $S(A)$ of a subregion $A$ of $\Sigma$, one can use the replica trick as follows:
\begin{align}
 S(A) &= \lim_{n\rightarrow1} \frac{\log\left( Z^{(b)}\left[M_{n}\right]\right) - n \log \left(Z^{(b)}\left[M_{1}\right]\right)}{1-n}
\nonumber\\
 &= \left(1-n\frac{d}{dn}\right) \log \left(Z^{(b)}[M_{n}]\right)\bigg{\vert{}}_{n\rightarrow1},
\end{align}
where
\begin{align}
 Z^{(b)}[M] = \int D\phi\, \exp\left(-\int_{M} d^{D}\!x \sqrt{\gamma}\, \mathcal{L}[\phi]\right)
\label{eq:EEdefn}
\end{align}
is the ``bare'' partition function computed by the path integral on a given manifold $M$. $M_{1}$ is the manifold used to compute ${\rm Tr}\, \rho$, while $M_{n}$ is an $n$-sheeted version of $M_{1}$ which is a branched cover with a conical excess of angle $\Delta\phi=2\pi(n-1)$ localized at the ($D-2$)-dimensional submanifold $\partial A$.

The bare partition function $Z^{(b)}[M]$ typically diverges and takes the form
\begin{align}
 \log \left(Z^{(b)}[M]\right) = c_{1} (\Lambda a)^{D} + c_{2} (\Lambda a)^{D-2} + \dots,
\label{eq:divergeZ}
\end{align}
where $\Lambda$ is a UV cutoff and $a$ is the length scale associated with the manifold $M$ \cite{komargodski2015aspects}. What are the contributions of these divergences to entanglement entropy? These divergence can be expressed as local integrals of background quantities \cite{Hung:2011ta,Liu:2012eea,Taylor:2016aoi}. (In even dimensions, there is a logarithmic divergence which cannot be expressed in this manner.) This implies that their contributions cancel in Eq.~(\ref{eq:EEdefn}) everywhere away from $\partial A$, since $M_{n}$ and $n$ copies of $M_{1}$ are identical manifolds except at $\partial A$. However, $M_{n}$ has extra divergent contributions coming from curvature invariants localized at $\partial A$. This leads to
\begin{align}
 S(A) = \sum_{k=1}^{\lfloor D/2\rfloor}a_{k}\Lambda^{D-2k}\! \int_{\partial A}\! d^{D-2}x \sqrt{H}  \left[\mathcal{R},\mathcal{K}^{2}\right]^{k-1},
\label{eq:EEdiverge}
\end{align}
where $[\mathcal{R},\mathcal{K}^{2}]^{k-1}$ represents all possible scalar intrinsic and extrinsic curvature invariants of $\partial A$ of mass dimension $2k-2$, with their coefficients collectively written as $a_{k}$, and $H_{ab}$ is the intrinsic metric of $\partial A$. Here, we have suppressed possible finite terms to focus on the leading divergences. This is the famous ``area law'' associated with entanglement entropy, which comes from the short distance correlations between $A$ and $\bar{A}$.

Since the above behavior is sensitive to the cutoff, one often considers a renormalized version of entropy. In particular, the divergences in Eq.~(\ref{eq:divergeZ}) can be subtracted (except for logarithmic ones) by introducing a counterterm action $I_{\text{ct}}$ which involves local integrals of curvature invariants:
\begin{align}
 I_{\text{ct}} = \sum_{k=1}^{\lfloor D/2 \rfloor +1} b_{k} \Lambda^{D-2k+2} \int_{M}d^{D}\!x \sqrt{\gamma}\, \mathcal{R}^{k-1}.
\label{eq:counter}
\end{align}
Here, $\mathcal{R}^{k-1}$ represents all possible scalar curvature invariants of $M$ that one can write down at mass dimension $2k-2$, and their coefficients $b_{k}$ can be tuned exactly to cancel the divergences. The renormalized entropy is then given by
\begin{align}
 S_{\text{ren}}(A) = \lim_{n\rightarrow1} \frac{\log \left(Z_{\text{ren}}[M_{n}]\right)-n\log \left(Z_{\text{ren}}[M_{1}]\right)}{1-n},
\label{eq:renEE}
\end{align}
where $Z_{\text{ren}}$ is the renormalized partition function which is computed using the action with the counterterms in Eq.~(\ref{eq:counter}). This renormalized entropy is universal, i.e.\ UV regulator independent in the continuum limit, and has been discussed previously in the literature \cite{Taylor:2016aoi}. A closely related version of renormalized entropy was discussed in Ref.~\cite{Liu:2012eea}. These two are not identical, but they both extract the appropriate universal behavior in the CFT limit by subtracting the power divergences.

\subsection{Entanglement Entropy in $TT$ Deformed Theories}
\label{sec:QFTcalc}

We now specialize to the case of a $D$-dimensional CFT deformed by
a particular composite operator $X_{D}$ of the stress tensor \cite{Hartman:2018tkw}. The presence of this deforming irrelevant operator breaks conformal invariance and gives rise to a QFT that is conjectured to be holographically dual to AdS with a finite cutoff radius, where Dirichlet boundary conditions are imposed.

We will focus on computing the partition function of this $TT$ deformed theory on the manifold $S^{D}$ of radius $R$:
\begin{align}
 \gamma_{ij} = R^2 d\Omega_D^2.
\label{eq:bdrymetric}
\end{align}
The theory is defined by the flow equation dictated by $X_D$, and using this we obtain
\begin{align}
 \langle T_{i}^{i}\rangle = -D\lambda\langle X_{D}\rangle,
\label{eq:TT}
\end{align}
where $\lambda$ is the deformation parameter. $T_{ij}$ is the renormalized stress tensor, whose trace vanishes up to conformal anomalies in the CFT limit $\lambda \rightarrow 0$. The bare stress tensor $T_{ij}^{(b)}$ is related to the renormalized one\footnote{The bare stress tensor is related to the Brown-York stress tensor \cite{Brown:1992br}, while the renormalized stress tensor is related to the Balasubramanian-Kraus stress tensor \cite{Balasubramanian:1999re} by a factor of $r_c^{d-2}$.} as
\begin{align}
 \langle T_{ij}^{(b)} \rangle = \langle T_{ij} \rangle - C_{ij},
\label{eq:bareT}
\end{align}
where $C_{ij}$ represent various terms involving the background metric $\gamma_{ij}$ that arise from variation of the counterterm action, which in the CFT limit is given by Eq.~(\ref{eq:counter}). For finite $\lambda$, the cutoff of the theory is provided by the deformation itself, so that $\Lambda$ is replaced by---or identified with---$\lambda^{-1/D}$ in Eqs.~(\ref{eq:divergeZ}~--~\ref{eq:counter}).

Since $S^{D}$ is a maximally symmetric space, the one point function
of the stress tensor takes the form
\begin{align}
 \langle T_{ij} \rangle       &= \omega_{D}(R)\,\, \gamma_{ij}, \\
 \langle T_{ij}^{(b)} \rangle &= \omega_{D}^{(b)}(R)\,\, \gamma_{ij}.
\label{eq:sphereT}
\end{align}
Using the flow equation, one can solve for $\omega_{D}(R)$ and $\omega_{D}^{(b)}(R)$ as has been done in Ref.~\cite{Caputa:2019pam}, yielding
\begin{align}
 \omega_{D}(R) =& -\frac{D-1}{2D\lambda}\sqrt{1+\frac{L_D^2}{R^{2}}}+ \frac{D-1}{2D\lambda} 
\nonumber\\
 & {} + \sum_{k=1}^{\lfloor(D-1)/2\rfloor} \frac{f_{k,D}}{\lambda}\left(\frac{L_D}{R}\right)^{2k},
\label{eq:omega}\\
 \omega_{D}^{(b)}(R) =& -\frac{D-1}{2D\lambda}\sqrt{1+\frac{L_D^2}{R^{2}}},
\label{eq:omega2}
\end{align}
where $L_D^2 = 2D(D-2)\alpha_{D}\lambda^{2/D}$ with $\alpha_D$ being quantities related to the central charges of the field theory, and $f_{k,D}$ are dimension dependent constants. (Note that $\alpha_D \propto 1/(D-2)$, so that $L_2 \neq 0$.) We stress that while $\omega_D(R)$ has been represented schematically, the expression for $\omega_D^{(b)}(R)$ is exact. The explicit expressions for $\omega_D(R)$ can be found in Ref.~\cite{Caputa:2019pam}.

Now using these results, we can compute the bare partition function as
\begin{align}
 \frac{d}{dR} \log Z^{(b)}_{S^D} = -\frac{1}{R}\int_{S^D} d^{D}\!x \sqrt{\gamma}\, \langle T_{i}^{i(b)} \rangle,
\end{align}
obtaining
\begin{align}
 \log Z^{(b)}_{S^D} &= - D\Omega_D\,\int_{0}^{R} dR\,\,\omega_{D}^{(b)}(R)\, R^{D-1} \nonumber\\
 &= \frac{\Omega_D L_D R^{D-1}}{2\lambda}  \,_{2}F_{1}\biggl[-\frac{1}{2},\frac{D-1}{2};\frac{D+1}{2};-\frac{R^2}{L_D^2}\biggr],
\end{align}
where $\Omega_D$ is the volume of a unit $S^D$. The entanglement entropy of a subregion $A$ which is a hemisphere of the spatial $S^{D-1}$ can then be computed by a simple trick described in Ref.~\cite{Donnelly:2018bef}:
\begin{align}
 S(A) &= \left(1-n\frac{d}{dn}\right) \log \left(Z^{(b)}[S^D_n]\right) \bigg{\vert}_{n\rightarrow1} \nonumber\\
 & = \left(1-\frac{R}{D}\frac{d}{dR}\right) \log Z^{(b)}_{S^D}.
\label{eq:spheretrick}
\end{align}
This gives us the answer
\begin{align}
 S(A) = \frac{\pi \Omega_{D-2} L_D R^{D-1}}{D(D-1) \lambda} \,_{2}F_{1}\biggl[\frac{1}{2},\frac{D-1}{2};\frac{D+1}{2};-\frac{R^{2}}{L_D^2}\biggr].
\label{eq:QFTEE}
\end{align}

We can also compute the renormalized entanglement entropy in multiple different ways, e.g.\ using Eq.~(\ref{eq:renEE}), which results in a universal answer in the CFT limit \cite{Taylor:2016aoi}. Alternately, one can use the version employed in Ref.~\cite{Liu:2012eea}. For finite $\lambda$ these two versions give different answers, which explains the discrepancy in Ref.~\cite{Banerjee:2019ewu} between the field theory calculation and the renormalized entropy.

We, however, emphasize that the $TT$ deformation provides a particular physical regulator for the entropy, so one need not focus their attention on the renormalized entropy. This regularization has a simple interpretation in field theory, which also has a geometric bulk interpretation. Specifically, on the field theory side one only includes the energy levels below the shock singularity, above which the energies take complex values. The existence of this regularization naturally leads us to consider the bare entanglement entropy in Eq.~(\ref{eq:QFTEE}), which captures all the information about correlations between $A$ and $\bar{A}$.

\section{Bulk Calculation}
\label{sec:bulk}

\subsection{Holographic Duality}
\label{sec:bulkbasic}

Using the holographic dictionary in Eq.~(\ref{eq:dict}), the entanglement entropy $S(A)$ of a boundary subregion $A$ can be calculated as
\begin{align}
 S(A) = \lim_{n\rightarrow1} \frac{I_{\text{bulk}}[B_{n}] - n I_{\text{bulk}}[B_{1}]}{n-1},
\label{eq:LM}
\end{align}
where $B_{n}$ and $B_{1}$ are the saddle point bulk solutions dual to the boundary conditions dictated by the field theory path integral on $M_{n}$ and $M_{1}$, respectively \cite{Lewkowycz:2013nqa}. Notably, the action $I_{\text{bulk}}$ dual to the bare partition function is the usual Einstein-Hilbert action supplemented by the Gibbons-Hawking-York boundary term. Assuming that the solution $B_n$ preserves the $\mathbb{Z}_n$ symmetry of the boundary, Ref.~\cite{Lewkowycz:2013nqa} showed that the contribution to the above expression is localized to the extremal surface $\Gamma$, resulting in the RT formula
\begin{align}
 S(A) = \min_{\partial \Gamma = \partial A} \left[\frac{\norm{\Gamma}}{4G}\right].
\label{eq:RTderive}
\end{align}
Our simple observation is that this proof carries through unmodified as long as one is computing the bare partition function. The $TT$ theory must then obey the RT formula by construction.

Counterterms added to the boundary action are well understood to correspond to boundary terms added to the bulk action \cite{Balasubramanian:1999re,Skenderis:2002wp}. Per the discussion in Section~\ref{sec:QFTbasic}, these terms give rise to extra contributions to $S(A)$ localized to the entangling surface $\partial A$. The saddle point solutions are not modified by the inclusion of these terms, which are pure functionals of the induced metric $h_{ij}$. This implies that the renormalized entropy can be calculated holographically as
\begin{align}
 S_{\text{ren}}(A) = \min_{\partial \Gamma = \partial A} \left[\frac{\norm{\Gamma}}{4G}\right] + \tilde{S}(\partial A),
\label{eq:holoRenEE}
\end{align}
where the form of $\tilde{S}(\partial A)$ is discussed in Ref.~\cite{Taylor:2016aoi}.

\subsection{RT Formula in Cutoff AdS}
\label{sec:bulkTT}

As a simple check, we now compare the result of the RT formula to the entanglement entropy obtained in Section~\ref{sec:QFTcalc}. On the bulk side, we need to compute the minimal surface $\Gamma$ anchored to the entangling surface $\partial A$ on the cutoff surface at $r = r_c$, on which the induced metric is given by
\begin{align}
 h_{ij} = r_c^2 R^2 d\Omega_D^2 \equiv r_0^2 d\Omega_D^2.
\end{align}
This calculation was performed in Ref.~\cite{Banerjee:2019ewu} and the answer obtained is
\begin{align}
 S(A) = \frac{r_0^{D-1} \Omega_{D-2}}{4G(D-1)} \,_{2}F_{1}\biggl[\frac{1}{2},\frac{D-1}{2};\frac{D+1}{2};-\frac{r_0^2}{l^2}\biggr],
\label{eq:HEEbulk}
\end{align}
where $l$ is the AdS radius. By using the holographic identifications
\begin{align}
 \lambda &=\frac{4\pi G l}{D r_c^D}, \\
 l^2 &= 2D(D-2)\alpha_D \lambda^{2/D} r_c^2 = L_D^2 r_c^2,
\end{align}
we find that this is identical to Eq.~(\ref{eq:QFTEE}).

\section{Discussion}
\label{sec:disc}

\subsection{Holographic Dictionary}

As emphasized throughout, if there exists a holographic duality between Einstein gravity in the bulk and a quantum field theory on the boundary such that the two are related by Eq.~(\ref{eq:dict}), then the RT formula will hold. This is true independent of the details of the bulk spacetime and the boundary field theory. Indeed, we have shown that the $TT$ deformed CFT provides an explicit example of the validity of the Lewkowycz-Maldacena (LM) proof beyond AdS/CFT at the conformal boundary.\footnote{An important assumption is the $\mathbb{Z}_n$ symmetry in the bulk. It is plausible that the argument holds after relaxing this assumption; See, e.g., Ref.~\cite{Camps:2014voa}.} In fact, all the results based only on this dictionary element will hold in any such duality, at least in a perturbative expansion in $G$. Two salient examples include the prescription for calculating refined R\'{e}nyi entropies presented in Ref.~\cite{Dong:2016fnf} and generalizations of the RT formula in higher curvature gravity \cite{Dong:2013qoa}. Though the robustness of the LM proof is far from unknown, we hope that highlighting this feature helps solidify the relationship between entanglement entropy and geometry in general spacetimes.

\subsection{Holographic Renormalization and Counterterms}

In CFT calculations, one often considers only renormalized quantities because these are universally well-defined and survive the continuum limit. However, entanglement entropy is not one of these quantities. Nevertheless, since the $TT$ operator implements a particular physical cutoff which has a simple geometric dual, it is sensible to consider bare quantities. In particular, $TT$ deformations with different background geometries would implement different regularizations, leading to different entanglement entropies. On the bulk side, this manifests as different choices of the cutoff surface. This provides a better understanding of the UV-IR correspondence.

The handling of counterterms is the only additional subtlety encountered when calculating entanglement entropy in $TT$ deformed CFTs. For finite deformations, all quantities are automatically regulated and hence the previous distinction between finite and divergent terms becomes muddled. We aimed to clarify the conceptual aspects of these terms and how they are related with the holographic result.

The fundamental idea is that the dictionary relation
\begin{align}
 Z_{\text{CFT}}[\gamma_{ij}] = e^{-I_{\text{bulk}}[g_{\mu\nu}]}
\label{eq:dict2}
\end{align}
is precisely between the \textit{bare} CFT on the boundary and Einstein-Hilbert gravity (plus the necessary Gibbons-Hawking-York term) in the bulk, both of which have divergent partition functions. This is the arena in which the RT formula was shown to hold. If one now chooses to introduce specific counterterms to renormalize the CFT stress tensor, then this will correspondingly alter the gravity side of the dictionary (specifically by adding terms localized to the boundary of the bulk). In particular, the addition of counterterms will alter the RT prescription to include terms beyond the standard area piece. This addition manifests as integrals of local geometric invariants at the entangling surface. In the CFT limit these are used to cancel power divergences, but with finite deformations one need not add a counterterm. Indeed, calculations including counterterms \cite{Donnelly:2018bef,Banerjee:2019ewu} would necessarily miss the area law piece for $D > 2$, which is finite for finite deformations.

%

\subsection{Holography in General Spacetimes}

The explicit verification of the RT formula beyond AdS/CFT at the conformal boundary of AdS provides a strong footing for the surface-state correspondence \cite{Miyaji:2015yva} and related constructions to understand holography in general spacetimes via entanglement entropy \cite{Sanches:2016sxy,Nomura:2016ikr,Nomura:2017npr,Nomura:2017fyh,Nomura:2018kji}. In previous work, the RT formula was used as an assumption to investigate properties of a hypothetical boundary theory and self consistency checks provided confidence in that assumption. Now, the evidence that a duality in the form of Eq.~(\ref{eq:dict}) exists beyond basic AdS/CFT, and the RT formula along with it, suggests that a duality may indeed exist for general spacetimes and bolsters our confidence in previous work.

The results of $TT$ deformations provide a particularly promising avenue to investigate flat space holography, since they hold down to scales below the AdS radius $l$. This contrasts with the conventional UV-IR correspondence, which would result in a single matrix-like theory describing an AdS volume \cite{Susskind:1998dq}. It suggests that there is a way to redistribute degrees of freedom on the boundary theory in a way that maintains local factorization, and the $TT$ deformation implements this. This is explicitly seen in the calculation of entanglement entropy in the fact that it does not face an obstruction when a volume law scaling is reached at $r_0 \approx l$. Volume law scaling of entanglement entropy suggests that the boundary theory for asymptotically flat space is non-local, as is expected from the $TT$ deformation. Corresponding behavior is seen in cosmological spacetimes \cite{Nomura:2017fyh}, and investigating properties of highly deformed CFTs may shed light on these theories. Since the $TT$ operator naturally implements some sort of coarse graining, it would be interesting to relate this to the geometric coarse graining procedure developed in Ref.~\cite{Nomura:2018kji}.

\acknowledgments

P.R. would like to thank Shouvik Datta and especially Vasudev Shyam for multiple discussions. This work was supported in part by the Department of Energy, Office of Science, Office of High Energy Physics under contract DE-AC02-05CH11231 and award DE-SC0019380, by the National Science Foundation under grant PHY-1521446,and by MEXT KAKENHI Grant Number 15H05895.

\bibliography{mybibliography}

\end{document}